\documentclass[aps,twocolumn,showpacs,amsmath,amssymb,floatfix]{revtex4}
\usepackage{amsthm,times}
\usepackage{latexsym}
\usepackage{amsfonts}
\usepackage{bbm,dsfont}
\usepackage{graphicx}

\newcommand{\C}{\mathbb{C}}
\newcommand{\Id}{\mathbb{I}}
\newcommand{\hil}{{H}}
\newcommand{\tr}[1]{\mathrm{Tr}\left[ {#1} \right]} 
\newcommand{\Tr}[2]{\mathrm{Tr}_{#1}\left[ {#2} \right]}

\newcommand{\ketbra}[2]{\left\vert {#1} \right\rangle\left\langle {#2} \right\vert}

\begin{document}
\title{
The canonical Naimark extension for generalized measurements
involving sets of Pauli quantum observables chosen at random}
\author{Carlo Sparaciari}
\affiliation{Dipartimento di Fisica, Universit\`a degli Studi di Milano, 
I-20133 Milano, Italy}
\author{Matteo G. A. Paris}
\email{matteo.paris@fisica.unimi.it}
\affiliation{Dipartimento di Fisica, Universit\`a degli Studi di Milano, 
I-20133 Milano, Italy}
\affiliation{CNISM, UdR Milano, I-20133 Milano, Italy}
\date{\today}
\begin{abstract}
We address measurement schemes where certain observables $X_k$ 
are chosen at random within a set of non-degenerate isospectral observables 
and then measured on repeated preparations of a physical system. 
Each observable has a probability $z_k$ to be measured, with 
$\sum_k z_k = 1$, and the statistics of this generalized 
measurement is described by a positive operator-valued measure (POVM).
This kind of schemes are referred to as {\em quantum roulettes} since each 
observable $X_k$ is chosen at random, e.g. 
according to the fluctuating value of an external parameter.
Here we focus on quantum roulettes for qubits involving the measurements of 
Pauli matrices and we explicitly evaluate their canonical 
Naimark extensions, i.e. their implementation as indirect measurements 
involving an interaction scheme with a probe system. We thus provide 
a concrete model to realize the roulette without destroying
the signal state, which can be measured again after the measurement, 
or can be transmitted. Finally, we apply our results to 
the description of Stern-Gerlach-like experiments on a two-level system.
\end{abstract} 
\maketitle
\section{Introduction}
In this paper we deal with a specific class of generalized quantum
measurements usually referred to as {\em Quantum Roulettes}. These
quantum measurements are achieved through the following procedure.
Consider $K$ projective measurements, described by the set $\{ X_k \}_{k
= 1, \ldots , K}$ of non-degenerate isospectral observables in a Hilbert
space $H$. The system is sent to a detector which, at random, performs
the measurement of the observable $X_k$.  Each observable has a
probability $z_k$ to be measured, with $\sum_k z_k = 1$. This scheme is
referred to as quantum roulette since the measured observable $X_k$ is
chosen at random, e.g. according to the fluctuating value of a physical
parameter, as it happens for the outcome of a roulette wheel. The
generalized observable actually measured by the detector is described by
a positive operator-valued measure (POVM), which provides the
probability distribution of the outcomes and the post-measurement
states \cite{Hel73,hel,hol}.
\par
As a matter of fact, any POVM on a given Hilbert space may be
implemented as a projective measurement in a larger one, e.g. see
\cite{Bin07} for single-photon qudits.  This
measurement scheme is usually referred to as a Naimark extension of the
POVM. Indeed, it is quite straightforward to find a Naimark extension
for the POVM of any quantum roulette in terms of a joint measurement
performed on the system under investigation and an ancillary one. 
\par
On the other hand, for any POVM the Naimark theorem \cite{NMK} ensures the
existence of a {\em canonical} Naimark extension, i.e. the
implementation of the POVM as an indirect measurement involving an
{\em independent} preparation of an ancillary (probe) system \cite{Per90}, an 
interaction of the probe with the system under investigation, and 
a final step where only the probe is subjected to a (projective) 
measurement \cite{bergou,epj}.  
A question thus arises on the
canonical implementation of the quantum roulette' POVM, and on the
resources needed to realize the corresponding interaction scheme. This
is the main point of this paper. In particular, we focus on quantum
roulettes involving the measurements of Pauli matrices on a qubit system
and explicitly evaluate their canonical Naimark extensions.
\par
We remind that having the Naimark extension of a generalized measurement 
is, in general, highly desirable, since it provides a concrete model 
to realize an apparatus which performs the measurement without 
destroying the state of the system under investigation. Thereby, the 
state after the measurement can be measured again, or can be 
transmitted, and the tradeoff between information gain and measurement
disturbance may be evaluated \cite{KB,OptFid,mista,cvqrep}.
Alternatively, the scheme may serve to perform indirect quantum control, 
\cite{iqc}. 
\par
It should be emphasized that the concept of quantum roulette provides a
natural framework to  describe measurement scheme where the measured
observable depends on an external parameter, which can not be fully
controlled and fluctuates according to a given probability distribution.
A prominent example is given by the Stern-Gerlach apparatus, which
allows one to measure a spin component of a particle in the direction
individuated by an inhomogeneous magnetic field \cite{am,qstg}. Indeed, whenever the
field is fluctuating, or the uncertainty in the splitting force
is taken into account \cite{Gar99}, 
the measurement scheme is described by a
quantum roulette. Also in this case, if we have the canonical Naimark
extension for the roulette, then we have a concrete way to realize a
measurement without destroying the state \cite{stg}.  We remind that in the
continuous variable regime quantum roulettes involving homodyne
detection with randomized phase of the local oscillator has been
already studied
theoretically \cite{qr} and realized experimentally \cite{munroe95}.
\par
The paper is structured as follows. In the next Section we introduce 
notation, briefly review the Naimark theorem, and gather all the
necessary tools, e.g. the Cartan decomposition of SU(4) transformations, 
which allows us to greatly reduce the number of parameters involved in
the problem of finding the canonical Naimark extension.
In Section \ref{s:qr} we introduce the concept of quantum roulette,
derive the corresponding POVM, and illustrate an example of 
non-canonical Naimark extension. In Section \ref{s:nmqr} we derive 
the canonical extension for one-parameter 
Pauli quantum roulettes and discuss details of their
implementation, whereas Section \ref{s:sg} is devoted to 
analyze Stern-Gerlach-like experiments as quantum roulettes, i.e. 
taking into account the possibility that the magnetic field is 
randomly fluctuating. Finally, Section \ref{s:out} closes the paper 
with some concluding remarks.
\section{Notation and tools}
\subsection{POVMs and the Naimark theorem}
When we measure an observable on a quantum system, we can not predict
which outcome we will obtain in each run. What we know is the spectrum of
possible outcomes and their probability distribution.
Given a system described by a state $\rho$ in the Hilbert space $\hil$, 
to obtain the probability distribution of the outcomes $x$ we use the Born Rule:
\begin{equation*}
p_x = \tr{\rho \Pi_x}
\end{equation*}
\noindent In order to satisfy the properties of the distribution $p_x$, the operators
$\Pi_x$ do not need to be projectors. The operators $\Pi_x$ have to be positive,
$\Pi_x \geq 0$, since the probability distribution
$p_x$ has to be positive for every $| \varphi \rangle \in \hil$, and
normalized, $\sum_x \Pi_x = \mathbb{I}$, since $p_x$ is normalized.
A decomposition of identity by positive operators $\Pi_x$
will be referred to as a {\it positive operator-valued measure} (POVM),
and the operators $\Pi_x$ are the elements of the POVM.
\par
We use $\Pi_x$ to get information about the probability distribution $p_x$,
but if we are interested in post-measurement states we have to introduce
the set of operators $M_x$, the {\it detection operators}.
These operators should give the same probability distribution given by $\Pi_x$,
thus they are obtained from $p_x = \tr{M_x \rho M_{x}^\dagger} = \tr{\rho \Pi_x}$.
Therefore, detection operators that satisfy $\Pi_x = M_x M_{x}^\dagger$ are
$M_x = U_x \sqrt{ \Pi_x }$, with $U_x$ a unitary operator such that $U_x U_{x}^\dagger = \Id$,
and this leaves a residual freedom on the post-measurement states.
The post-measurement states are then given by:
\begin{equation*}
\rho_x = \frac{1}{p_x} M_x \rho M_x^\dagger
\end{equation*}
\noindent A measurement described by the operators $\Pi_x$ is referred to as
{\it generalized measurement}.
\par
In order to link general measurements with physical schemes of measurement,
we have the Naimark theorem, which states that a generalized measurement in a Hilbert space $\hil_A$
may be always seen as an indirect measurement in a larger Hilbert space given by
the tensor product $\hil_A \otimes \hil_B$. This indirect measure is known as
{\it canonical Naimark extension} for the generalized measurement.
Conversely, when we perform a projective measure on the subsystem $\hil_B$
of a composite system $\hil_A \otimes \hil_B$, the degrees of freedom of $\hil_B$
may be traced out and we obtain the same probability distribution of the outcomes
of the projective measurement and the same post-measurement states
performing a generalized measurement on the subsystem $\hil_A$.
\begin{figure}[h!]
\centering
\includegraphics[width=0.4\columnwidth]{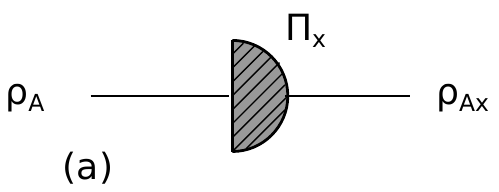}
\includegraphics[width=0.52\columnwidth]{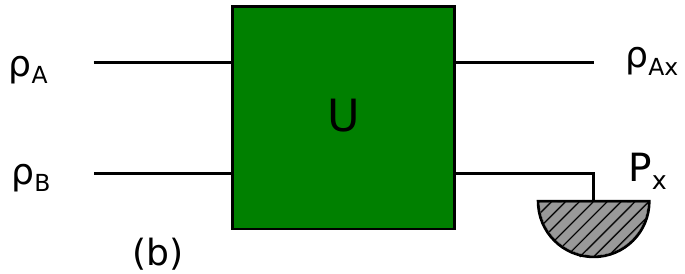}
\caption{The two measurement schemes linked by the Naimark theorem.
(a): a generalized measurement described by the POVM $\Pi_x=M^\dag_x
M_x$; (b): its canonical Naimark extension, defined by the triple
$\{ \rho_B, U , \{ P_x \} \}$, describing the probe state $\rho_B$,
the evolution operator $U$ and the projective measurement $\{P_x\}$
on the probe system, respectively.}
\label{figPOVM}
\end{figure}
\par
The Naimark theorem gives a practical recipe to evaluate the canonical
extension for a generalized measurement in a Hilbert space $\hil_A$:
\begin{equation} \label{extension} \Pi_x = \Tr{B}{\Id \otimes \rho_B\;
U^\dagger\; \Id \otimes P_x\; U} \end{equation} \noindent where $\rho_B
\in L(\hil_B)$ describes the state of the probe system (or {\it
ancilla}), the operators $\{P_x\} \in L(\hil_B)$ are a set of projectors
which describe the measurement on the ancilla and the unitary operator
$U \in L(\hil_A \otimes \hil_B)$ works on both the system and the
ancilla.  A canonical Naimark extension for the generalized measurement
given by the operators $\{\Pi_x\} \in L(\hil_A)$ is thus individuated by
the triple $\{ \rho_B , U , \{ P_x \} \}$.
\par
Evaluating the canonical Naimark extension for a generalized measurement is
desirable since it gives a concrete model to realize an apparatus which performs
the measurement without destroying the state.
Then, the post-measurement state can be transmitted, or measured again.
\subsection{The Cartan decomposition of $SU(4)$ transformations}
\begin{figure}[h!]
    \centering
    \includegraphics[width=0.95\columnwidth]{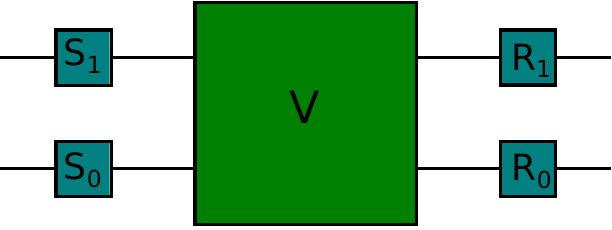}
    \caption{The Cartan decomposition of the operator 
      $X \in SU(4)$, given by $( R_1 \otimes R_0 ) V ( S_1 \otimes S_0)$.}
    \label{figCartan}
\end{figure}
In the following we are going to deal with two-qubit interactions, 
i.e. unitary operators (with unit determinant) of the group $SU(4)$, 
which are individuated by 15 parameters. In order to reduce the number 
of these parameters we will make use of the Cartan decomposition, which
allows us to factor a general operator in $SU(4)$ into local operators 
working on single qubits and a single two-qubit operator $V$
individuated by 3 parameters \cite{c1,c2,c3}, see (Fig. \ref{figCartan}).
\par
According to the Cartan decomposition any $X \in SU(4)$ can be rewritten as 
$X = ( R_1 \otimes R_0 ) V ( S_1 \otimes S_0)$, where $R_1 , R_0 , S_1 , S_0 \in SU(2)$ 
and $V = \mathrm{exp} \{ i \sum_{j=1}^3 k_j \sigma_j \otimes \sigma_j \}$,
with $\mathbf{k} \equiv (k_1 , k_2 , k_3) \in \mathbb{R}^3$.
The operators $\sigma_i$ are the Pauli matrices.
If we introduce the following equivalence relation in $SU(4)$
\begin{equation} \label{equivalence}
A \sim B \quad if \quad A = ( R_1 \otimes R_0 ) B ( S_1 \otimes S_0 )
\end{equation}
we can split the whole space of unitary operators with 
unit determinant into equivalence classes. Then, since an operator $X \in SU(4)$
is represented (by the equivalent relation given here) by the matrix $V$,
it is possible to establish a link between operators in $SU(4)$ and real vectors:
$X \sim \mathbf{k}$, where $\mathbf{k}$ is the {\it class vector} of $X$.
The following operations are class-preserving:
\begin{description}
\item[Shift:] $\mathbf{k}$ can be shifted by $\pm \frac{\pi}{2}$ along one of its components.
\item[Reverse:] the sign of two components of $\mathbf{k}$ can be reversed.
\item[Swap:] two components of $\mathbf{k}$ can be swapped.
\end{description} 
By the use of these operations it is always possible to 
reduce any $\mathbf{k}$ into a bounded region $K$ given by:
\begin{enumerate}
\item $\frac{\pi}{2} > k_1 \geq k_2 \geq k_3 \geq 0$
\item $k_1 + k_2 \leq \frac{\pi}{2}$
\item $\mathrm{If}\ k_3 = 0 ,\ \mathrm{then}\ k_1 \leq \frac{\pi}{4}$
\end{enumerate}
The $\mathbf{k} \in K$ are referred to as {\it canonical class vectors}.
\par
The expression of the operator V may be simplified using a different set
of parameters, e.g.
\begin{align*}
k_1 &= - \frac{\alpha_1 - \alpha_2}{4} \\
k_2 &= - \frac{\alpha_1 + \alpha_2}{4} \\
k_3 &= - \frac{\alpha_3}{2}
\end{align*}
Besides, we introduce the operators $\Sigma_i = \frac12 \sigma_i \otimes
\sigma_i$, which are normalized in the space of $4 \times
4$ operators, with the inner product
$\langle A , B \rangle = \mathrm{Tr}[ B^\dagger A ]$. 
Eventually, we obtain the following matrix $V$:
\begin{equation} 
\label{matrixV}
V = \mathrm{exp}\left\{-i\left[ 
\frac12 (\alpha_1 - \alpha_2) \Sigma_1 + 
\frac12 (\alpha_1 + \alpha_2) \Sigma_2 +
\alpha_3 \Sigma_3 \right]\right\}
\end{equation}
In terms of the $\alpha$ parameters the bounded region corresponding to
the canonical class vectors is given by 
\begin{align*}
-\pi \leq \alpha_1 & \leq 0\\
0 \leq  \alpha_2 & \leq -\alpha_1\\
 \alpha_1 + \alpha_2 \leq  2 \alpha_3 & \leq 0\\
\hbox{if } \alpha_3 = 0 \hbox{ then } \alpha_1 - \alpha_2 & \geq -\pi
\end{align*}
As we will see in the next section, the Cartan decomposition simplifies 
the problem of finding the canonical Naimark extension for the quantum 
roulette since we will be able to neglect single-qubit operations
and thus reducing the the 15-parameters operator $U$ to the 
3-parameters operator $V$. Furthermore, we will restrict the interval 
of the parameters of $V$ to the bound region defined above. 
\section{The Pauli quantum roulette wheel}
\label{s:qr}
Let us consider $K$ observables $\{ X_k \}$ in a Hilbert space
$\hil_A$ with dimension $d = \mathrm{dim}(\hil_A)$.
All the observables are non-degenerate and isospectral.
Since the observables are non-degenerate and the Hilbert space is finite-dimensional,
each of them has $d$ eigenvalues.
We use a detector which chooses at random one of these observables and
performs a measurement of that observable.
Each observable has a probability $z_k$ of being selected by the detector
and $\sum_k z_k = 1$. This scheme, denoted by the $K$-tuple
$\{ \{ X_1 , z_1 \} , \{ X_2 , z_2 \} , \ldots , \{ X_K , z_K \} \}$,
is referred to as {\it Quantum Roulette}.
\par
If we have a system represented by the state $\rho \in L(\hil_A)$
and we send it to the detector, the probability distribution of the 
outcomes is given by
\begin{align} \label{roulette}
p_x &= \sum_k z_k p_{x}^{(k)} = \sum_k z_k \tr{\rho P_x^{(k)}}\\ \nonumber
&= \tr{\rho\,\sum_k z_k P_x^{(k)}} = \tr{\rho\, \Pi_x}
\end{align}
\noindent where $p_{x}^{(k)}$ is the probability distribution of the outcome $x$
for the observable $X_k$ and $P_x^{(k)} = | x \rangle^{(k)(k)}\langle x |$ is
the 1-dimensional projector on the eigenspace of the eigenvalue $x$
for the observable $X_k$. In the last equality of the Eq.\ (\ref{roulette})
we have introduced the POVM of the roulette, whose elements are given by
\begin{equation} \label{POVMroulette}
\Pi_x = \sum_k z_k P_x^{(k)} 
\end{equation}
\par
The $\Pi_x$'s are positive operators, indeed, given any $| \varphi \rangle \in \hil_A$, we have
$\langle \varphi | \Pi_x | \varphi \rangle = \sum_k z_k |\langle \varphi | x \rangle^{(k)}|^2 \geq 0$,
and they represent a decomposition of identity, since
\begin{equation*}
\sum_x \Pi_x = \sum_x \sum_k z_k P_x^{(k)} = \sum_k z_k \sum_x P_x^{(k)} = \sum_k z_k\; \mathbb{I} = \mathbb{I}
\end{equation*}
\noindent On the other hand, $\Pi_x$ are not orthogonal projectors since
$\Pi_x \Pi_{x'} \neq \delta_{xx'}\; \Pi_x$. Indeed:
\begin{equation*}
\Pi_x \Pi_{x'} = \sum_{k,k'} z_k z_{k'} P_x^{(k)} P_{x'}^{(k')} \neq 0
\end{equation*}
\noindent In fact, while ${}^{(k)}\langle x | x' \rangle^{(k)}$ has to be equal to $\delta_{xx'}$
for a fixed value of $k$, the quantity ${}^{(k)}\langle x | x' \rangle^{(k')}$
(with $k \neq k'$) could be different from zero also when $x \neq x'$.
\subsection{The non-canonical Naimark extension}
The quantum roulette has a Naimark extension that is not the canonical one
(i.e.\ the indirect measurement scheme described by the Naimark theorem)
that can be obtained as follow:
consider an additional Hilbert space $\hil_B$, describing the ancilla,
with dimension equal to the number $K$ of observables $X_k \in L(\hil_A)$,
and a basis $\{ | \theta_k \rangle \}$ in $\hil_B$.
Then we introduce projectors $Q_x = \sum_k P_x^{(k)} \otimes | \theta_k \rangle \langle \theta_k |$
in the larger Hilbert space $\hil_A \otimes \hil_B$ 
and we prepare the initial state of the ancilla in
$| \omega_B \rangle = \sum_k \sqrt{z_k} | \theta_k \rangle$,
obtaining the probability distribution
\begin{equation*}
p_x = \mathrm{Tr}_{AB}[ \rho \otimes | \omega_B \rangle \langle \omega_B |\; Q_x ]
\end{equation*}
\noindent that gives us the POVM's elements
\begin{equation*}
\Pi_x = \mathrm{Tr}_B[ \mathbb{I} \otimes | \omega_B \rangle \langle \omega_B |\; Q_x ] = \sum_k z_k P_x^{(k)}
\end{equation*}
\par
Moreover, the post-measurement states will be given by
\begin{align*}
\rho_x &= \frac{1}{p_x} \mathrm{Tr}_B[ Q_x\; \rho \otimes | \omega_B \rangle \langle \omega_B |\; Q_x ] \\
&= \frac{1}{p_x} \sum_k z_k\; P_x^{(k)} \rho P_x^{(k)}
\end{align*}
\noindent This measurement scheme does not involve an evolution operator
$U \in L(\hil_A \otimes \hil_B)$ and the projective measure
is performed on both system and ancilla,
unlike the canonical extension that involves a projective measurement on the sole ancilla. 
\section{The canonical Naimark extension of the Pauli quantum roulette
wheel}
\label{s:nmqr}
We focus on quantum roulettes which work on qubit systems with Hilbert 
space $\hil_A \equiv \C^2$ and address quantum roulettes involving the 
measurement of Pauli operators. In order to obtain the canonical Naimark extension
for these roulettes we have to add a probe system (the ancilla). We
assume a two-dimensional ancilla and show that this is enough to realize
the canonical extension using Eq. (\ref{extension}),
where the elements of the POVM are given by Eq. (\ref{POVMroulette}).
\par
Since $\hil_B$ is a Hilbert space with dimension two, we choose the
following representation for the state $\rho_B = \ketbra{\omega_B}{\omega_B}$
and for the projector $P_x = \ketbra{x}{x}$, where
$| \omega_B \rangle = \cos \frac{\theta}{2} | 0 \rangle + e^{i \varphi} \sin \frac{\theta}{2} | 1 \rangle$
and $| x \rangle = \cos \frac{\alpha}{2} | 0 \rangle + e^{i \beta} \sin \frac{\alpha}{2} | 1 \rangle$,
with the parameters $\alpha , \theta \in [ 0 ; \pi ]$
and $\beta , \varphi \in [ 0 ; 2 \pi )$.
Notice that $| 0 \rangle , | 1 \rangle$ are a basis in $\hil_B$;
we assume that $| 0 \rangle$ is the eigenvector of $\sigma_3$ related to the eigenvalue 1,
while $| 1 \rangle$ is the eigenvector related to -1.
\par
The last tool to individuate the canonical extension is the evolution operator
$U \in L(\hil_A \otimes \hil_B)$ which works on the overall state of the composite system.
The operator $U \in SU(4)$, then it is defined by 15 parameters.
Therefore, the total number of parameters that defines the Naimark extension is 19
(4 parameters from $| \omega_B \rangle$ and $| x \rangle$ plus 15 from $U$).
As we will see this number can be greatly reduced by employing the Cartan decomposition.
\subsection{Application of the Cartan decomposition to the Naimark extension}
The Naimark theorem provides a practical connection between the generalized measurement
given by the quantum roulette and the indirect measurement described by the extension.
Indeed both the probability distribution of the outcomes $p_x$ and the post-measurement
states $\rho_{Ax}$ have to be equal for these two schemes. That is:
\begin{equation} \label{born_conn}
p_x = \mathrm{Tr}_{AB}[ U \rho_A \otimes \rho_B U^\dagger \mathbb{I} \otimes P_x ] = \mathrm{Tr}_A[ \rho_A \Pi_x ]
\end{equation}
\begin{equation} \label{reduction_conn}
\rho_{Ax} = \frac{1}{p_x} \mathrm{Tr}_B[ U \rho_A \otimes \rho_B U^\dagger \mathbb{I} \otimes P_x ] = \frac{1}{p_x} M_x \rho_A M_x^\dagger
\end{equation}
\noindent where the distribution $p_x$ and the state $\rho_{Ax}$ in the first
equality belong to the projective measurement while
those in the last equality belong to the generalized measurement.
\par
We focus now on the Born rule Eq.\ (\ref{born_conn}) in order to evaluate the operators $\Pi_x$;
after straightforward calculation, we obtain the elements of the POVM:
\begin{equation*}
\Pi_x = S_1^\dagger\; \mathrm{Tr}_B[ ( \mathbb{I} \otimes S_0 \rho_B S_0^\dagger ) V^\dagger ( \mathbb{I} \otimes R_0^\dagger P_x R_0 ) V ]\; S_1
\end{equation*}
\par
Consider now $S_0 \rho_B S_0^\dagger$ and $R_0^\dagger P_x R_0$;
the operators $R_0 , S_0$ $\in L(\hil_B)$ represent a rotation in the qubit Hilbert space $\hil_B$.
Since both $\rho_B$ and $P_x$ are not yet defined and depend on some parameters,
we can combine the rotation to them and we are left with a transformation from $L(\hil_B)$ to $L(\hil_B)$:
\begin{equation*}
\rho_B \rightarrow \rho_B' = S_0 \rho_B S_0^\dagger
\end{equation*}
\begin{equation*}
P_x \rightarrow P_x' = R_0^\dagger P_x R_0
\end{equation*}
\noindent i.e.\ we can neglect this transformation by
a suitable reparametrization of $\rho_B'$ and $P_x'$.
Furthermore, we assume the operator $S_1$ to be the identity ($S_1 = \mathbb{I}$).
We make this assumption in order to simplify the research of the canonical extension.
This ansatz will be justified a posteriori:
once we find the Naimark extension, if the probability distribution $p_x$ obtained from the extension
is equal to the one obtained from the POVM, then the extension is correct and $S_1 = \mathbb{I}$.
\par
Now we have POVM's elements obtained by the canonical extension:
\begin{equation} \label{real_extension}
\Pi_x = \mathrm{Tr}_B[ ( \mathbb{I} \otimes \rho_B ) V^\dagger ( \mathbb{I} \otimes P_x ) V ]
\end{equation}
\noindent and these elements have to be equal to those evaluated for the quantum roulette in exam.
Using the Cartan decomposition on the canonical Naimark extension reduces the number of parameter
to 7 (4 from $\rho_B$ and $P_x$ and 3 from $V$).
\subsection{Detection operators for the Quantum Roulette}
The operator $R_1$ is not involved in the definition of the elements of the POVM,
but it is necessary for the evaluation of the post-measurement state $\rho_{Ax}$.
Since the operators $R_0$ and $S_0$ were absorbed into, respectively,
$P_x$ and $\rho_B$ and $S_1 = \mathbb{I}$,
then the decomposition of $U$ is $U = (R_1 \otimes \mathbb{I}) V$ and
the left part of Eq.\ (\ref{reduction_conn}) becomes:
\begin{align*}
\rho_{Ax} &= \frac{1}{p_x} \mathrm{Tr}_B[ (R_1 \otimes \mathbb{I}) V \rho_A \otimes \rho_B V^\dagger ( R_1^\dagger \otimes P_x ) ] \\
&= \frac{1}{p_x} R_1 \mathrm{Tr}_B[ V \rho_A \otimes \rho_B V^\dagger ( \mathbb{I} \otimes P_x ) ] R_1^\dagger
\end{align*}
\par
Therefore, the operator $R_1$ describes a residual degree of freedom
in the design of possible post-measurement states.
This freedom was expected, since when we define a POVM $\Pi_x$,
the post-measurement states can be evaluated using the detection operators $M_x$.
These operators are defined as $M_x = U_x \sqrt{\Pi_x}$,
where $U_x$ is a unitary operator.
The operator $U_x$ provides the same freedom given by $R_1$ to the post-measurement states.
\subsection{The general solution}
The problem of finding the canonical Naimark extension for
a given quantum roulette is now basically reduced to the solution
of four equations dependent on seven parameters.
Indeed, the considered roulettes are always in qubit spaces;
hence, the elements of the POVM $\Pi_x$ are self-adjoint $2 \times 2$ operators
on the field $\mathbb{C}$ and the relation given by Eq.\ (\ref{real_extension})
provides four equations: one from the element $\Pi_{x11}$ that is real,
two from the element $\Pi_{x12}$ (real part and imaginary part)
and one from the element $\Pi_{x22}$.
\subsection{Exchange of the parameters}
One may wonder if it is possible to look for the canonical extension
when the parameters $\alpha_i$ get values from all $\mathbb{R}$.
The Cartan decomposition does not impose restriction on the range
of the components of the class vectors (that is, the parameters $\alpha_i$),
but we know that each operator in $SU(4)$ is related
(via the equivalence relation Eq.\ (\ref{equivalence}))
to a canonical class vector, whose components lie on the bounded region $K \in \mathbb{R}^3$.
On the other hand if we find an extension with $\alpha_i$ outside of $K$,
it is possible to use the 3 class-preserving operations
(shift, reverse and swap) to bring back the parameters to $K$.
\par
May the parameters be brought back to $K$ after we have found the canonical extension?
This is not possible, unless we also modify the other objects of the extension.
Indeed, if we have found the extension, then we have defined both
$\alpha_i$ and $\theta,\varphi,\alpha,\beta$.
But if the $\alpha_i$ are modified by one of the three class-preserving operations,
then also the other parameters are modified and the state
$| \omega_B \rangle$ and the operator $P_x$ change.
In fact, the operations are class-preserving, so they transform the operator $V$ into
\begin{equation*}
V \rightarrow  ( R_1 \otimes R_0 ) V' ( S_1 \otimes S_0)
\end{equation*}
\noindent and the operators $R_0 , S_0 \in SU(2)$ modify both the initial state of $H_B$ and the orthogonal projector
\begin{align*}
\rho_B(\theta',\varphi') = S_0\ \rho_B(\theta,\varphi)\ S_0^\dagger \\
P_x(\alpha',\beta') = R_0^\dagger\ P_x(\alpha,\beta)\ R_0
\end{align*}
\noindent Hence, to transform the $\alpha_i$ and keep the correct extension is necessary to modify $\rho_B$ and $P_x$.
\subsection{The canonical extension}
We now focus on quantum roulettes given by Pauli operators $\sigma_i$ and
on their canonical Naimark extension. The most general quantum roulette
of this kind is $\{ \sigma_i , z_i \}_{i=1,2,3}$ and  its canonical extension
depends on two undefined parameters (e.g.\ $z_1$ and $z_2$). Finding 
the extension for the general roulette is analytically challenging
and thus we focus to roulettes involving two Pauli operators.
\par
Let us consider the roulette $\{ \{ \sigma_1 , z \} , \{ \sigma_3 , 1 - z \} \}$,
where $z$ gets values from the interval $(0;1)$. The POVM's
elements are give by
\begin{equation}
\Pi_1 = \frac{1}{2}
\begin{pmatrix}
2 - z & z \\
z & z
\end{pmatrix} \quad 
\ \Pi_{-1} = \frac{1}{2}
\begin{pmatrix}
z & -z \\
-z & 2 - z
\end{pmatrix}
\end{equation}
and the detection operators by $M_x = U_x \sqrt{\Pi_x}$, $x=\pm 1$.
Upon expanding them on the Pauli basis, i.e. 
$M_x = a_0\, \mathbb{I} + a_1\,\sigma_1 + a_2\, \sigma_2 + a_3\,
\sigma_3$, the coefficients $a_i$ are evaluated using the inner product
$\langle X , Y \rangle = \mathrm{Tr}[X Y^\dagger]$.
For the roulette $\{ \{ \sigma_1 , z \} , \{ \sigma_3 , 1 - z \} \}$,
we obtain $a_i = a_i(z)$ (for $i = 0,1,3$) and $a_2 = 0$.
In other words, the detection operators of a roulette involving the 
Pauli operators $\sigma_1$ and $\sigma_3$ have no component on 
the missing one, i.e. $\sigma_2$. This result also holds for 
the other roulettes depending on two $\sigma$'s , e.g. 
for $\{ \{ \sigma_2 , z \} , \{ \sigma_3 , 1 - z \} \}$, 
the detection operators $M_x$ have no component by $\sigma_1$. 
\par
The solution for the canonical extension corresponds to the parameters
\begin{align*}
\alpha_1 &= - \pi\ ;\ \alpha_2 = 0\ ;\ \alpha_3 = \arcsin ( - \sqrt{ 
\frac{1}{1 - \frac{z}{2} } -1 } ) \\
\alpha &= \arccos (z - 1 )\ ;\ \beta = \pi\ ;\ \theta = 
\frac{\pi}{2}\ ;\ \forall \varphi\,,
\end{align*}
where $\alpha_1 , \alpha_2 , \theta , \varphi$ and $\beta$
are in the correct range and we are left 
to check whether also $\alpha_3$ and $\alpha$ lies in the correct range.
First of all, $\cos \alpha = z - 1$, i.e.\ $\cos \alpha \in (-1 ; 0 )$;
then $\alpha \in ( \frac{\pi}{2} ; \pi )$.
Finally, $\sin \alpha_3 = - \sqrt{ \frac{1}{1 - \frac{z}{2} } -1 }$,
that is $\sin \alpha_3 \in (-1 ; 0)$; then $\alpha_3 \in ( -\frac{\pi}{2} ; 0 )$.
The parameter $\alpha_3$ has to be in $[ \frac{ \alpha_1 + \alpha_2 }{2} ; 0 ]$,
and since $\alpha_1 = - \pi$ and $\alpha_2 = 0$, its greatest range is $[- \frac{\pi}{2} ; 0]$.
\par
The ingredients of the canonical extension are thus the state 
$| \omega_B \rangle = \frac{1}{ \sqrt{2} } | 0 \rangle + 
\frac{e^{i \varphi}}{ \sqrt{2} } | 1 \rangle$, the projectors 
\begin{align}
P_1 &= \frac{1}{2}
\begin{pmatrix}
2 - z & \sqrt{ z ( 2 - z)} \\
\sqrt{ z ( 2 - z)} & z
\end{pmatrix} \\
P_{-1} &= {\mathbb I} - P_1 \nonumber
\end{align}
and the unitary 
\begin{equation}
V =
\begin{pmatrix}
f(z) & 0 & 0 & 0 \\
0 & 0 & i\ f^\ast(z) & 0 \\
0 & i\ f^\ast(z) & 0 & 0 \\
0 & 0 & 0 & f(z)
\end{pmatrix}
\end{equation}
\noindent with $f(z) = \sqrt{ \sqrt{ \frac{2 - 2 z}{2 - z} } + \frac{i}{ \sqrt{ \frac{2}{z} -1} } }$.
\par
In order to obtain the canonical Naimark extension for the roulettes
$\{ \{ \sigma_1 , z \} , \{ \sigma_2 , 1 - z \} \}$ and
$\{ \{ \sigma_2 , z \} , \{ \sigma_3 , 1 - z \} \}$,
we have to remove our previous assumption $S_1 = \Id$.
In fact, to rotate a Pauli operator $\sigma_i$ by an angle $\theta$ we have to use
a rotation operator $W = e^{-i ( \mathbf{n} \cdot \mathbf{ \sigma } ) \theta}$,
where $\mathbf{n}$ is the versor of the direction around which the rotation is made.
Then, to move from a two Pauli operators roulette to another,
we need to apply the correct rotation in order to modify the $\sigma_i$.
For example, to move from $\{ \{ \sigma_1 , z \} , \{ \sigma_3 , 1 - z \} \}$
to $\{ \{ \sigma_2 , z \} , \{ \sigma_3 , 1 - z \} \}$ we have to apply
the operator $W = e^{-i \frac{\pi}{4} \sigma_3}$,
which changes $\sigma_1$ into $\sigma_2$ and leaves $\sigma_3$ unchanged.
\par
Therefore, the extensions for the other roulettes depending on two Pauli operators
are defined by the same parameters of the extension for $\{ \{ \sigma_1 , z \} , \{ \sigma_3 , 1 - z \} \}$,
but the elements $\Pi_x$ are rotated by the operator $W$, that is:
\begin{equation*}
\Pi_x \rightarrow W \Pi_x W^\dagger
\end{equation*}
\noindent This means that, while for the first found extension the operator
$S_1$ can be assumed equal to $\mathbb{I}$,
for the extensions of the other roulettes the operator $S_1$ has to be equal
to the conjugate transpose of the rotation operator $W$.
We find that, for the roulette given by $\sigma_2$ and $\sigma_3$, $S_1 = e^{i \frac{\pi}{4} \sigma_3}$
while for the one given by $\sigma_1$ and $\sigma_2$, $S_1 = e^{-i \frac{\pi}{4} \sigma_1}$.
\section{The Stern-Gerlach apparatus as a quantum roulette}
\label{s:sg}
The so called Stern-Gerlach apparatus allows one to measure a component
(e.g.the component along the $z$-axis $S_z$) of the quantum observable
{\it spin}, i.e.\ the intrinsic angular momentum of a particle.  The
measurement is usually performed on a collimated beam of particles
(e.g.\ neutral atoms) sent with thermal speed into a region of
inhomogeneous magnetic field.  Here the particles are deflected by the
field in some beams which, after propagating into the vacuum, are
collected by a screen. The magnetic field is usually assumed of 
the form $\mathbf{B} = (B - b z )\, \mathbf{e}_3$, where $z$ is the
cohordinate along the $z$-axis, $B$ is the field in the origin, and 
$b$ is a constant. Actually, this is an artificial model, since a 
field like this one does not respect the Maxwell equations, as 
$\mathbf{\nabla} 
\cdot  \mathbf{B} = -b \neq 0$. On the other hand, we may assume 
that $b \ll B$ so that we can neglect the other components 
of $\mathbf{B}$.
The interaction Hamiltonian is given by $H =  (B - b z) 
\sigma_3$ (neglecting the vacuum permittivity) and the 
corresponding evolution operator by $U = e^{-i \tau  ( B - b z) \sigma_3}$, 
where $\tau$ is an effective interaction time.
Starting from an initial state which is factorized into a spin and a  
spatial part, i.e. $| \Psi \rangle\!\rangle = 
(c_0 |0\rangle + c_1 |1\rangle) \otimes
|\psi (\mathbf{q})\rangle$, the evolved state is given by
$$
U | \Psi \rangle\!\rangle = 
c_0 |0\rangle \otimes |\psi_- (\mathbf{q})\rangle + c_{1} |1\rangle \otimes
|\psi_+(\mathbf{q})\rangle
$$
where $|\psi_{\pm} (\mathbf{q}) \rangle = e^{\pm i \tau  ( B - b z)} 
| \psi (\mathbf{q}) \rangle$. The evolution is thus coupling the spin
and the spatial part. Moving to the momentum representation 
\begin{align*}
| \widetilde{\psi}_\pm (\mathbf{p}) \rangle &= 
\int\!d^3 \mathbf{q}\,
e^{-i \mathbf{q} \cdot \mathbf{p}\,} | \psi_\pm (\mathbf{q})\rangle
\\ &= |\widetilde{\psi} ( \mathbf{p} \pm \tau b \mathbf{e}_3) \rangle
\,
\end{align*}
and tracing out the spin part after the interaction, 
we have that the motional degree of freedom after the interaction is 
described by the density operator 
\begin{align*}
\varrho_{\mathbf{p}} = & 
|c_1|^2 | 
\widetilde{\psi}_+ (\mathbf{p})
\rangle\langle
\widetilde{\psi}_+ (\mathbf{p})|
+
|c_0|^2 | 
\widetilde{\psi}_- (\mathbf{p})
\rangle\langle
\widetilde{\psi}_- (\mathbf{p})|
\end{align*}
As a consequence the beam is divided in two parts, and it is possible to 
perform measurements on a screen placed at a given distance from the 
magnetic  field, where we can see the beams as two different spots.
\par
If for some reason the direction of the magnetic field is tilted 
we have $\mathbf{ B } = (B - b t )\,\mathbf{e}_\alpha$, where $t$ is a
cohordinate along the new direction and $\mathbf{e}_\alpha=\cos\alpha
\,\mathbf{e}_3 +\sin\alpha\, \mathbf{e}_\perp$, $\mathbf{e}_\perp$ denoting any
direction perperdicular to the $z$-axis, say $\mathbf{e}_1$. 
The above analysis is still valid if we perform the substitution
$H \rightarrow  (B - b t) \sigma_\theta$ where
$\sigma_\theta$ is a Pauli matrix describing a sping component along
a tilted axis. Assuming a rotation along the $x$-axis we have 
that $\sigma_\theta$ corresponds to the rotated operator 
\begin{equation*}
\sigma_\theta  = U_\theta \sigma_3 U^\dagger_\theta
\end{equation*}
where $U_\theta  = e^{-i \sigma_1 \theta}$ and $\theta=\alpha/2$.
\subsection{The continuous quantum roulette}
Usually, the magnetic field of the apparatus is assumed to a be a stable 
classical quantity. However, in any practical situation the magnetic field 
unavoidably fluctuates. In particular, we focus on Stern-Gerlach
apparatuses where the magnetic field fluctuates in one dimension around a 
pre-established direction and
provide a more detailed analysis of non-ideal setups \cite{stg1,stg2}.  
As mentioned above, a measurement of spin made with a
tilted magnetic field corresponds to measure the operator $\sigma_\theta$. 
If the magnetic field is fluctuating, then we may describe
this situation using a continuous quantum roulette where $\theta$ is 
randomly fluctuating around the $z$-axis according to a given probability
distribution.
\par
In principle, the magnetic field may fluctuate in any direction
on the zy-plane, i.e. the angle $\theta$ takes values between 
$-\frac{\pi}{2}$ and $\frac{\pi}{2}$. On the other hand, in a realistic 
situation, the magnetic field moves away from
the pre-established direction (the $z$-axis, in this case) just by a small angle.
We thus introduce a Gaussian probability distribution $z(\theta)$ for
the fluctuating values of $\theta $
\begin{equation*}
z(\theta) = \frac{1}{A} \exp\{- \frac{\theta^2}{2 \Delta^2}\}
\end{equation*}
where the normalization A is:
\begin{equation*}
A = \int_{-\frac{\pi}{2}}^{\frac{\pi}{2}} 
\exp\{- \frac{\theta^2}{2 \Delta^2}\} 
d\theta = \sqrt{2 \pi} \Delta \mathrm{Erf}( \frac{\pi}{2 \sqrt{2} \Delta} )
\end{equation*}
In order to evaluate the elements of the POVM which describes
this continuous quantum roulette, we need the projectors on the
eigenspaces of $\sigma_\theta$, i.e. 
\begin{align}
P_1(\theta) &=
\begin{pmatrix}
\cos^2 \theta & i \cos \theta \sin \theta \\
-i \cos \theta \sin \theta & \sin^2 \theta
\end{pmatrix} \\ 
P_{-1}(\theta) & = {\mathbb I} - P_1(\theta) \nonumber
\end{align}
\noindent Therefore, the elements of the POVM are given by
\begin{equation} \label{cont_roulette}
\Pi_x = \int_{-\frac{\pi}{2}}^{\frac{\pi}{2}} z(\theta) P_x(\theta) d\theta
\end{equation}
\noindent that is the equation equivalent to Eq.\ (\ref{POVMroulette})
in the continuous case. We can evaluate the elements of the POVM using
the distribution $z(\theta)$ and the projectors $P_x(\theta)$ given above,
and we obtain:
\begin{align}
\Pi_1 &=
\begin{pmatrix}
\frac{1}{2} + f(\Delta) & 0 \\
0 & \frac{1}{2} - f(\Delta)
\end{pmatrix} \\
\Pi_{-1} &= {\mathbb I} - \Pi_1 \nonumber
\end{align}
where the function $f(\Delta)\ :\ [0 ; + \infty ) \rightarrow
[\frac{1}{2} ; 0 )$ is given by 
\begin{equation*}
f(\Delta) = \frac{ \mathrm{Erf}(\frac{\pi - i 4 \Delta^2}{2 \sqrt{2} \Delta}) 
+ \mathrm{Erf}(\frac{\pi + i 4 \Delta^2}{2 \sqrt{2} \Delta})}{4 e^{2 \Delta^2} 
\mathrm{Erf}(\frac{\pi}{2 \sqrt{2} \Delta})}\,.
\end{equation*}
We have $f(\Delta) \simeq 1/2 - \Delta^2$ for vanishing $\Delta$ and
$f(\Delta) \simeq 1/8\Delta^2$ for $\Delta \rightarrow \infty$.
\subsection{The canonical extension for the continuous roulette}
We look for the canonical extension for this roulette
in order to obtain a practicable measurement scheme with the same behavior
(same probability distribution and post-measurement states) of the Stern-Gerlach experiment
with fluctuating magnetic field. A canonical extension may be found,
corresponding to the parameters
\begin{align*}
\alpha_1 &= \arccos( -2 f(\Delta) )\ ;\ \alpha_2 = \arccos(2 f(\Delta))\ ;\ \alpha_3 = 0 \\
\alpha &= \pi\ ;\ \beta = 0\ ;\ \theta = 0\ ;\ \varphi = 0
\end{align*}
\par
Let consider the parameters $\alpha_1$ and $\alpha_2$;
the codomain of the function $f(\Delta)$ is $(0 ; \frac{1}{2}]$.
Therefore, if $\cos \alpha_1 = -2 f(\Delta)$,
then $\cos \alpha_1 \in (0 ; -1]$ and $\alpha_1 \in (-\frac{\pi}{2} ; -\pi]$.
Instead, $\cos \alpha_2 = 2 f(\Delta)$, i.e.\ $\cos \alpha_2 \in (0 ; 1]$ and
$\alpha_2 \in (\frac{\pi}{2} ; 0]$.
Both $\alpha_1$ and $\alpha_2$ depend on the function $f$,
so when we choose a value for $f$ the two parameters are fixed.
For example when $f(\Delta) \to 0$, then $\alpha_1 \to -\frac{\pi}{2}$
and $\alpha_2 \to \frac{\pi}{2}$; on the other hand, if $f(\Delta) = \frac{1}{2}$
then $\alpha_1 = -\pi$ and $\alpha_2 = 0$.
The ranges of these two parameters are correct and we have
$\alpha_2 \leq -\alpha_1\ \forall\ f(\Delta)$.
Finally, since we have fixed $\alpha_3 = 0$, we have to check 
whether $\alpha_1 - \alpha_2 \geq -\pi$, and this is case:
as it can be easily checked $\alpha_1 - 
\alpha_2 = -\pi$ for all $\Delta \in [0 ; + \infty )$. 
The canonical extension is thus given by the state 
$| \omega_B \rangle = | 0 \rangle$, the observable
$\sigma_3$ (measured on the ancilla), and the unitary $V \in L(H_A \otimes H_B)$
\begin{align}
V&=\frac14 \sum_{k=0}^3 v_k\, \sigma_k \otimes \sigma_k \\
v_{0/3} &= \sqrt{\frac12 + f(\Delta)}  \pm \sqrt{\frac12 - f(\Delta)} \\
&\stackrel{\Delta \rightarrow 1}{\simeq} 1 \pm \Delta \notag \\ 
v_{1/2} &= i\, v_{0/3}
\end{align}
For a particle with spin up,
represented by the pure state $| 0 \rangle$,
the probability distribution of the outcomes is given by 
$p_1 = \frac{1}{2} + f(\Delta)$,
$p_{-1} = 1- p_1$ and thus, when such a particle
is measured, there is always a probability 
that the apparatus measures the spin down $| 1 \rangle$.
\section{Conclusions}
\label{s:out}
We have addressed Pauli quantum roulettes  and found
their canonical Naimark extensions. The extensions are minimal, i.e 
they involve a single ancilla qubit, and provide 
a concrete model to realize the roulettes without destroying
the signal state, which can be measured again after the measurement, 
or can be transmitted. Our results provide a natural framework to 
describe measurement scheme where the measured observable depends 
on an external parameter, which can not be fully
controlled and may fluctuate according to a given 
probability distribution. As an illustrative example we
have applied our results to the description of Stern-Gerlach-like 
experiments on a two-level system, taking into account 
possible uncertainties in the splitting force.
\section*{Acknowledgments}
This work has been supported by MIUR through the project 
FIRB-RBFR10YQ3H-LiCHIS.

\end{document}